\documentclass[pre,showpacs,showkeys,preprintnumbers,amsmath,amssymb,superscriptaddress,twocolumn]{revtex4}
\pdfoutput=1 
\usepackage{graphicx}
\usepackage{amsfonts}
\usepackage{url}

\usepackage{epstopdf}
\usepackage{nameref,hyperref}
\usepackage{comment}
\usepackage{amsmath} 
\usepackage{lastpage,fancyhdr,graphicx}
\usepackage{color}

\def\det{{\textrm{det}}}

\def\M{{{\mathcal M}}}
\def\T{{{\mathcal T}}}
\def\Z{{{\mathbb Z}}}

\begin{document}

\title{Heterogeneous continuous time random walks}

\author{Denis~S.~Grebenkov}
 \email{denis.grebenkov@polytechnique.edu}
\affiliation{
Laboratoire de Physique de la Mati\`{e}re Condens\'{e}e (UMR 7643), \\ 
CNRS -- Ecole Polytechnique, 91128 Palaiseau, France}

\affiliation{Interdisciplinary Scientific Center Poncelet (ISCP),%
\footnote{International Joint Research Unit -- UMI 2615 CNRS/ IUM/ IITP RAS/ Steklov MI RAS/ Skoltech/ HSE, Moscow, Russian Federation} \\
Bolshoy Vlasyevskiy Pereulok 11, 119002 Moscow, Russia}

\author{Liubov Tupikina}
\affiliation{
Laboratoire de Physique de la Mati\`{e}re Condens\'{e}e (UMR 7643), \\ 
CNRS -- Ecole Polytechnique, 91128 Palaiseau, France}

\date{\today}

\begin{abstract}
We introduce a heterogeneous continuous time random walk (HCTRW) model
as a versatile analytical formalism for studying and modeling
diffusion processes in heterogeneous structures, such as porous or
disordered media, multiscale or crowded environments, weighted graphs
or networks.  We derive the exact form of the propagator and
investigate the effects of spatio-temporal heterogeneities onto the
diffusive dynamics via the spectral properties of the generalized
transition matrix.  In particular, we show how the distribution of
first passage times changes due to local and global heterogeneities of
the medium.  The HCTRW formalism offers a unified mathematical
language to address various diffusion-reaction problems, with numerous
applications in material sciences, physics, chemistry, biology, and
social sciences.
\end{abstract}





\maketitle

\section{Introduction}

Understanding transport phenomena in multiscale porous media and
crowded environments is of paramount importance in material sciences
(e.g. hardening of concretes or degradation of monuments caused by
salts penetration into stones), in petrol industry (oil extraction
from sedimentary rocks), in agriculture (moisture propagation in
soils), in ecology (contamination of underground water reservoirs and
streams), in chemistry (diffusion of reactants towards porous
catalysts), in biology (transport inside cells and organs, such as
lungs, kidney, placenta), to name but a few
\cite{Bouchaud,Sahimi1993,Coppens99,Kirchner00,Berkowitz2000,Song00,Berkowitz02,%
Scher02,Plassais05,Gallos07,Dentz08,Zoia09,Scher10,Berkowitz10,Bressloff2013,Metzler2014,Serov16}.
In spite of a significant progress in imaging techniques and
computational tools over the last decade, accurate modeling of these
processes is still restricted to a relatively narrow range of time and
length scales.  At the same time, the multiscale structure of porous
media has a critical impact onto the transport properties
\cite{Sahimi2012,Levitz2002,Levitz2012}.  For instance, concretes
exhibit pore sizes from few nanometers in the cement paste to few
centimeters (or larger) that greatly impacts water diffusion, the
consequent cement hydration and, ultimately, the mechanical properties
of the material.  Bridging theories and simulations on different
scales has become at the heart of modern approaches to such multiscale
phenomena.  In particular, one aims at coarse-graining an immense
amount of microscopic geometrical information about the medium from
high-resolution imaging, and revealing the structural features that
are the most relevant for a macroscopic description of the transport
processes.

\begin{figure}
\begin{center}
\includegraphics[width=45mm]{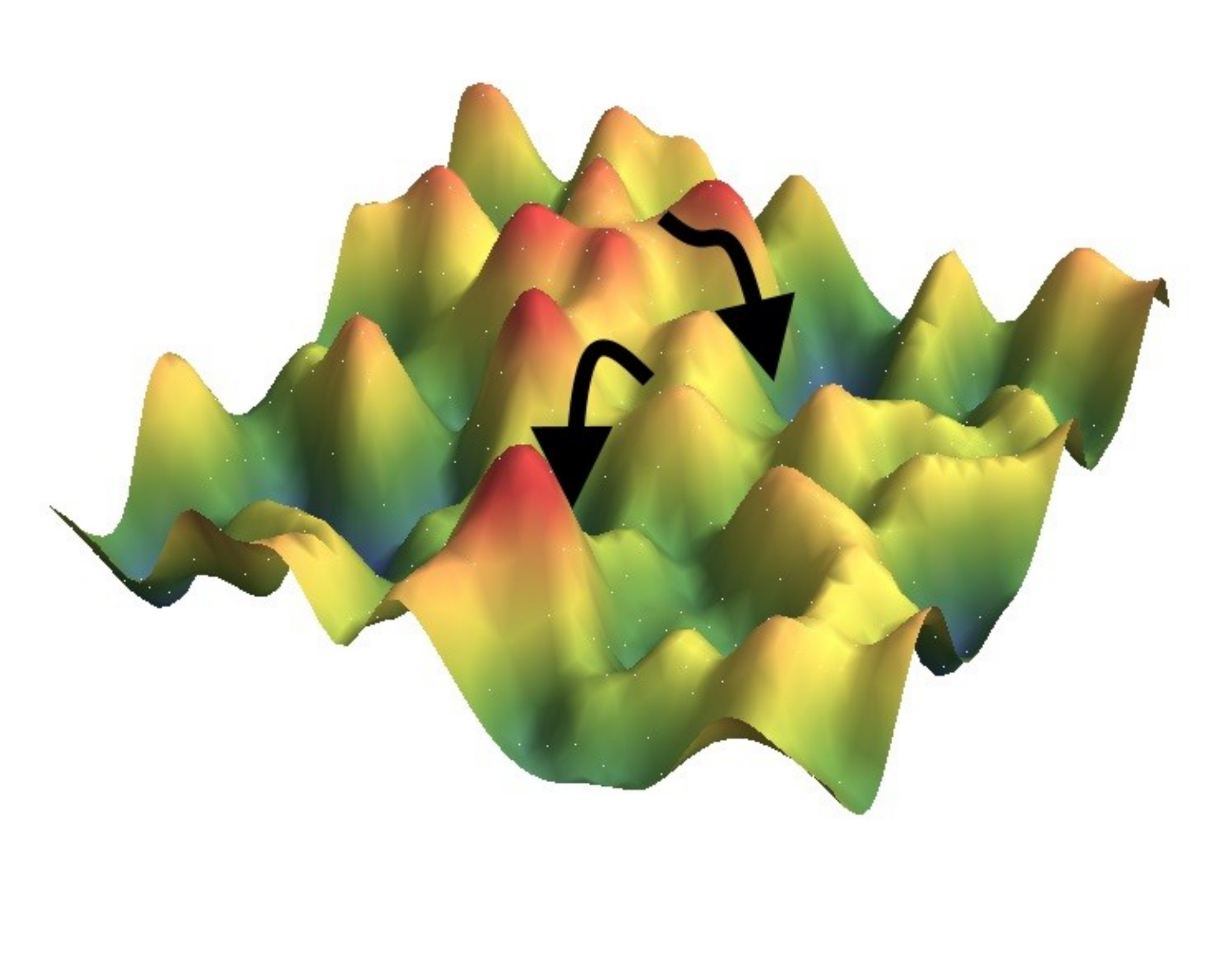}  
\includegraphics[width=40mm]{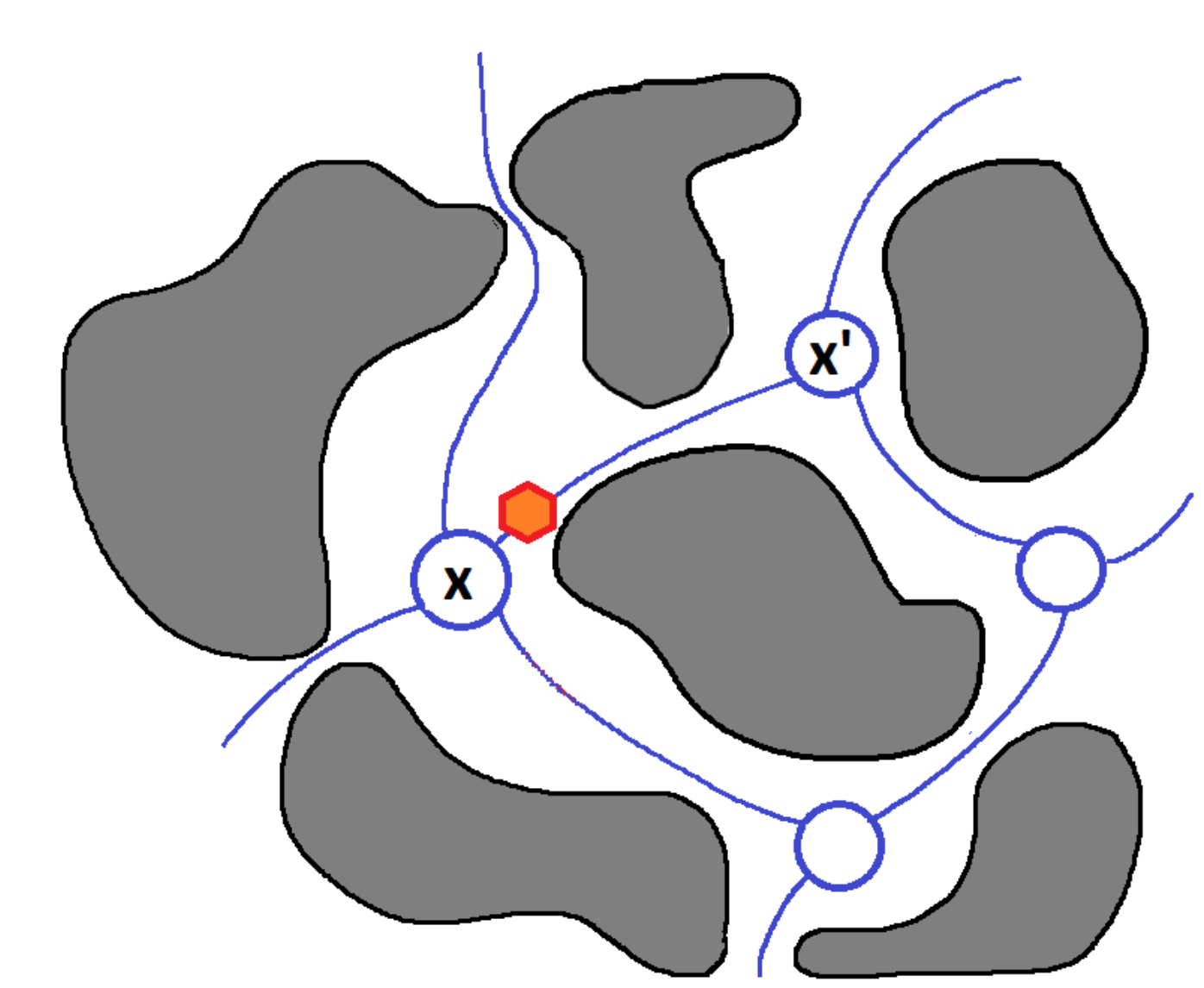}  
\end{center}
\caption{
(Left).  A complex dynamics in a disordered potential energy landscape
can be approximated as HCTRW between local minima (defining the sites
of the coarse-graining graph) with random exchange times $\T_{xx'}$
between neighboring sites $x$ and $x'$ drawn from an exponential
probability density $\psi_{xx'}(t)$ with the mean time $\tau_{xx'}
\propto \exp(U_{xx'}/kT)$, where $U_{xx'}$ is the energetic barrier
between two minima and $kT$ is the thermal energy.  (Right).
Diffusion of a particle (shown by a hexagon) inside a porous medium
(white space with gray obstacles) can be approximated as HCTRW between
pores (defining the sites of the coarse-graining graph) with random
travel times $\T_{xx'}$ between neighboring sites $x$ and $x'$ drawn
from an exit time (or travel time) probability density $\psi_{xx'}(t)$
determined by the shape of the pore at $x$ and its connections to
neighboring pores $x'$.}
\label{HCTRW_landscape}
\end{figure}

In this light, continuous time random walks (CTRWs), introduced by
Montroll and Weiss \cite{montroll1965,Montroll1969,Montroll1973}, have
been often evoked as an important model of diffusive transport in
disordered and porous media
\cite{Metzler1998,Metzler04,Barkai2008,Sahimi2012,Kang11,Fouxon2016,Amitai2017}.
In this model, a diffusing particle spends a random time at a region
of space (e.g., a pore) or at a site of a lattice before jumping to
another region or site.  The waiting event reflects either energetic
trapping of the walker in a local minimum of the potential energy
landscape, or a geometric trapping in a pore separated from other
pores by narrow channels (Fig.~\ref{HCTRW_landscape})
\cite{Bouchaud,Scher1973,Levitz2002,Levitz2012,Kenkre1973,Bouten1986}.  This
model is intrinsically homogeneous, as all sites have the same waiting
time distribution $\psi(t)$.  In practice, however, pores and channels
have a broad distribution of sizes and shapes, as well as the local
minima of the potential energy landscape are broadly distributed.  An
extension of the conventional approach by considering a site-dependent
waiting time distribution, $\psi_x(t)$, may capture the heterogeneity
of the minima or pore shapes but ignores heterogeneities in mutual
minima arrangements or in inter-pore connections.  For this reason, we
propose a more general approach that we call \emph{heterogeneous
continuous time random walk} (HCTRW).  In this approach, a random
walker moves on a graph, jumping from a site $x$ to a site $x'$ with
the probability $Q_{xx'}$.
The graph can be either a natural representation of the studied system
(e.g., electric, transportation, internet or social network), or
constructed as a coarse-grained representation of a potential energy
landscape or a porous medium (Fig.~\ref{HCTRW_landscape}).  Graphs can
also serve as discrete approximations (meshes) to Euclidean domains
and manifolds.  The \emph{travel (or exchange)} time
$\mathcal{T}_{xx'}$ needed to move from $x$ to $x'$ is a random
variable drawn from the probability density $\psi_{xx'}(t)$, which
depends on both sites $x$ and $x'$.  In this paper, we only consider
the Markovian case with independent jumps on connected graphs.

The paper is organized as follows.  In
Sec. \ref{hctrw_nonhetero_graph} we introduce the HCTRW framework and
derive the exact formula for the Laplace-transformed propagator.  We
show that the dynamics of HCTRW is fully determined by the spectral
properties of the generalized transition matrix that couples temporal
and spatial heterogeneities.  Using perturbation theory we establish
the long-time asymptotic behavior of the propagator for both finite
and infinite mean travel times.  In Sec. \ref{Discus_sec}, we show the
relation of the HCTRW formalism to multi-state switching models.  We
also discuss a natural inclusion of boundary conditions into the model
and the consequent possibility to assess various first passage
quantities and reaction kinetics in a unified way.  In particular, the
peculiar effects of spatio-temporal heterogeneities onto the first
passage time (FPT) distribution are presented.  An explicit solution
for the HCTRW propagator on $m$-circular graphs and some technical
derivations are reported in Appendices.

\section{General formalism}
\label{hctrw_nonhetero_graph}

\subsection{Propagator}	 
\label{anal_form_hctrw}

We derive the general formula for the propagator $P_{x_0x}(t)$ of the
HCTRW on a graph.  Here we adapt the matrix notation, writing $x_0$
and $x$ as subscripts.  The propagator $P_{x_0x}(t)$ is the
probability to find a walker at a site $x$ at time $t$ if it started
from a site $x_0$ at time $0$.  This probability can be written as
\begin{align}
P_{x_0x}(t)= \sum_{n=0}^{\infty}P_{x_0x}^{(n)}(t) ,
\label{propag_form}
\end{align}
where $P_{x_0x}^{(n)}(t)$ is the probability to find the random
walker, started from $x_0$, at $x$ at time $t$ after $n$ independent
jumps.  Note that the order of starting and ending points is important
since we consider a general, not necessarily symmetric, transition
matrix $Q$.  Each component $P_{x_0x}^{(n)}(t)$ can be represented as
\begin{align}
P_{x_0x}^{(n)}(t) = \int_{0}^{t}R_{x_0x}^{(n)}(t')\Psi_x(t-t')dt',
\label{p_n_t_x_0}
\end{align} 
where $\Psi_x(t-t')$ is the probability of staying at site $x$ during
time $t-t'$ and $R_{x_0x}^{(n)}(t')$ is the probability density to
reach $x$ from $x_0$ at time $t'$ at the $n^{th}$ step, which due to
the Markovian property is
\begin{align}
R_{x_0x}^{(n)}(t) = \int_{0}^{t} dt' \sum_{x'}R_{x_0x'}^{(n)}(t') Q_{x'x}(t-t') ,
\label{r_t_x_x0}
\end{align}
where $Q_{x'x}(t) $ is the joint transition probability density:
\begin{align}
Q_{x'x}(t) = Q_{x'x} \psi_{x'x}(t) .
\label{p_t_xx'}
\end{align}
The structural heterogeneities of the graph, represented by the
transition matrix $Q$
\cite{spitzer,Kehr1987,Weiss94,mugnolo2008,Sokolov2006,Angstmann13},
are now coupled, via the generalized transition matrix $Q(t)$, to
dynamical heterogeneities represented by the densities
$\psi_{x'x}(t)$.  In contrast to the Montroll-Weiss formula for
ordinary CTRW with a continuous jump distribution, there is no Fourier
transform in Eq.~(\ref{p_n_t_x_0}) because the probability
$P^{(n)}_{x'x}(t-t') $ is written for a discrete graph (see Appendix
\ref{sec:Montroll}).  Applying the Laplace transform to
Eq.~(\ref{p_n_t_x_0}) and using its linearity and convolution
property, one gets
\begin{align}
\tilde{R}_{x_0x}^{(n)}(s) = \sum_{x'}
\tilde{R}_{x_0x'}^{(n)}(s)  [\tilde{Q}(s)]_{x'x} ,
\label{lapl_trans_r}
\end{align}
where $\tilde{Q}(s)$ is the Laplace transform of $Q(t)$
\begin{align}
\tilde{Q}_{x'x}(s) = Q_{x'x}\tilde{\psi}_{x'x}(s),
\label{q_matr}
\end{align}
with the Laplace transform of quantities denoted with the tilde above
them, e.g., $\tilde{\psi}_{x'x}(s) = \int_{0}^{\infty} \psi_{x'x}(t)
e^{-st}dt$.  With this notation, we can write Eq.~(\ref{lapl_trans_r})
in a compact form
\begin{align}
\tilde{R}^{(n)}_{x_0x}(s) = [\tilde{Q}(s)^n]_{x_0x} .
\label{r_s_n}
\end{align}
Hence we get the Laplace transform of the propagator
$\tilde{P}_{x_0x}(s)$ using Eq.~(\ref{propag_form}):
\begin{equation}
\tilde{P}_{x_0x}(s) = [(I-\tilde{Q}(s))^{-1}]_{x_0x} \, \tilde{\Psi}_x(s) ,
\label{prop_x_0_x}
\end{equation}
where the geometric series formula was applied to the sum of powers
$(\tilde{Q}(s))^n$ given that $\| \tilde{Q}\| \leq 1$ (see Appendix
\ref{sec:Qnorm}).
Writing 
\begin{align}
\tilde{\Psi}_x (s) = \frac{1-\sum_{x'} \tilde{Q}_{xx'}(s)}{s} ,
\end{align}
the final expression of the propagator of HCTRW in the Laplace domain
is
\begin{align}
\tilde{P}_{x_0x}(s) = \frac{1-\sum_{x'}\tilde{Q}_{xx'}(s)}{s} 
[(I-\tilde{Q}(s))^{-1}]_{x_0x}.
\label{hctrw_final}
\end{align}
This is one of the main results of the paper.  Note that the
propagator determines all the moments of the position of the walker,
including the mean squared displacement.

The inverse Laplace transform is then needed to get the propagator in
time domain.  When the exchange times are drawn from exponential
distributions, $\tilde{\psi}_{xx'}(s) = (1 + s \tau_{xx'})^{-1}$,
$\tilde{P}_{x_0x}(s)$ in Eq. (\ref{hctrw_final}) is a ratio of two
polynomials of $s$, whereas $P_{x_0x}(t)$ gets the usual form of a sum
of exponentially decaying functions.  In this practically relevant
case, one needs to find the poles of $\tilde{P}_{x_0x}(s)$, i.e., the
zeros of the equation $\det(I-\tilde{Q}(s)) = 0$.  The Gerschgorin
theorem determines the radius of a disk in the complex plane, in which
the poles are located, and hence speeds up their numerical calculation
\cite{Carstensen1989}.  In the homogeneous case,
$\tilde{\psi}_{xx'}(s) = \tilde{\psi}(s) = (1 + s\tau)^{-1}$, the
problem is reduced to computing the eigenvalues $\lambda_k^0$ of the
matrix $H_0 = I - Q$ and then finding $s$ at which $\tilde{\psi}(s) =
1/(1-\lambda_k^0)$.  One gets thus the poles $s_k = -
\lambda_k^0/\tau$, as expected.  In general, however, spatio-temporal
heterogeneities in $\tilde{\psi}_{xx'}(s)$ can significantly alter the
above relation between the dynamical properties of the HCTRW
(determined by the poles $s_k$) and the spectral properties of the
stochastic matrix (the eigenvalues $\lambda_k^0$).  Moreover, if some
$\tilde{\psi}_{xx'}(s)$ are non-analytic, the Laplace-transformed
propagator can also be non-analytic.  As a consequence, $P_{x_0x}(t)$
may not be expressed as a sum of exponentials, exhibiting a slower
approach to the steady-state limit (see below).

\subsection{Spectral analysis and the long-time behavior}
\label{hctrw_finit}

In general, the matrix $H(s)= I-\tilde{Q}(s)$ is real but not
symmetric so that its complex-valued eigenvalues form complex
conjugate pairs \cite{Trefethen}.  For each $s$, we denote $u_k, v_k$
the left and right eigenvectors of $H(s)$, associated with the same
eigenvalue $\lambda_k$:
\begin{align}
u_k \, H(s) = {\lambda_{k}} \, u_{k}, \qquad
H(s) \,v_{k}= \lambda_{k} \, v_{k}.
\label{Hs_uk_vk}
\end{align} 
The left eigenvectors of $H(s)$ are just the transpose of the right
eigenvectors of the transposed matrix $H(s)^\dagger$.  One gets thus
the spectral representation of Eq.~(\ref{hctrw_final})
\begin{align}
\tilde{P}_{x_0x}(s)= 
\frac{1-\sum_{x'}\tilde{Q}_{xx'}(s)}{s} \sum_{k\geq 0} \frac{v_{k}(x_0) \, u_{k}(x)}{\lambda_{k}}  ,
\label{ps_spectral}
\end{align}
where we used bi-orthogonality: $(u_j \cdot v_k) = \delta_{j,k}$.
Since $u_{k}$ is a left row-vector, we do not write the transpose
symbol $\dagger$ for $u_{k}$.  Although the explicit dependence on
$x_0$ and $x$ is factored out in Eq. (\ref{ps_spectral}), this
representation remains rather formal, since $u_k, v_k$ and $\lambda_k$
depend on $s$ in a highly non-trivial way.  However it shows that the
spectral properties of the generalized transition matrix
$\tilde{Q}(s)$ fully determine the propagator of HCTRW.  In some
particular cases, the eigenvalues and eigenvectors of the matrix $H$
can be found explicitly, allowing one to derive an explicit form of
the propagator in time domain, as illustrated in Appendix
\ref{hctrw_circ} for $m$-circular graphs.  In general, however, the
time dependence of the propagator over the whole range of times is
difficult to grasp, and one focuses on long-time asymptotic behavior.

The long-time behavior of HCTRW is determined by $\tilde{P}_{x_0x}(s)$
at small $s$.  Here we distinguish two cases: (i) when all mean travel
times $\langle \T_{xx'}\rangle$ are finite, and (ii) when at least one
of the mean travel times is infinite.

In the former case, one gets the expansion
\begin{align}
\tilde{\psi}_{xx'}(s) = 1-s \langle \mathcal{T}_{xx'}\rangle + o(s).  
\end{align}
Introducing a matrix $T$ with elements 
\begin{align}  \label{eq:T}
T_{xx'} = Q_{xx'}\langle \mathcal{T}_{xx'}\rangle ,  
\end{align}
one gets $H(s) \approx I-Q+sT + o(s)$ so that the Laplace transform of
the propagator can be approximated as
\begin{align}
\tilde{P}_{x_0x}(s)\simeq 
t_x \bigg[(I-Q + sT)^{-1}\bigg]_{x_0x},
\label{prop_special_case}
\end{align}
where
\begin{align} 
t_x = \sum_{x'}T_{xx'} .
\label{eq:tx}
\end{align}
The normalization of the propagator is preserved even in this
approximate form (see Appendix \ref{normal_hctrw}).

Using the standard perturbation analysis at small $s$
\cite{Kato,Sternheim}, we substitute the expansions
\begin{subequations}
\begin{eqnarray}
\lambda_k &=& \lambda_{k}^0 +  s\lambda_{k}^1 + o(s),  \\ 
u_k &=& u_{k}^0 + s u_{k}^1 + o(s),  \\ 
v_k &=& v_{k}^0 + s v_{k}^1 + o(s)   
\end{eqnarray}
\end{subequations}
into Eq.~(\ref{Hs_uk_vk}) to get in the zeroth and first order in $s$:
\begin{subequations}
\begin{align}
&u_{k}^0 H_0 = \lambda_{k}^0 u_{k}^0, \quad
H_0 v_{k}^0 = \lambda_{k}^0 v_{k}^0, \\ 
&u_{k}^0 H_0 + u_{k}^0 T  = \lambda_{k}^0 u_{k}^0 + \lambda_{k}^1 u_{k}^0, \\ 
&H_0 v_{k}^0 + T v_{k}^0 = \lambda_{k}^0 v_{k}^0 + \lambda_{k}^1 v_{k}^0.
\end{align}
\end{subequations}
Multiplying the second equation by $u_{k}^0$, we get
\begin{align}
\lambda_{k}^1 = (u_{k}^0 T v_{k}^0).
\end{align}
Then to the first order Eq.~(\ref{ps_spectral}) becomes: 
\begin{align}
\tilde{P}_{x_0x}(s) \simeq t_x \sum_{k\geq 0} \frac{v_{k}^0(x_0) \, u_{k}^0(x)}{\lambda_{k}^0 + s\lambda_{k}^1}  .
\label{ps_first}
\end{align}
Given that $\lambda_{0}^0 = 0$ and $v_{0}^0 = 1/\sqrt{N}$ due to the
normalization of the transition matrix $Q$, where $N$ is the number of
vertices in the graph, it is convenient to isolate the term with $k =
0$:
\begin{align}
\tilde{P}_{x_0x}(s) \simeq \frac{p^{\rm st}_x}{s} + t_x 
\sum_{k>0} \frac{v_{k}^0(x_0) \, u_{k}^0(x)}{\lambda_{k}^0 + s\lambda_{k}^1} ,
\label{ps_first2}
\end{align}
where
\begin{align}
p^{\rm st}_x = \frac{t_x \pi_x}{\sum_{x'} t_{x'} \pi_{x'}}
\label{eq_stat}
\end{align}
is the steady-state (stationary) distribution, with $\pi_x$ being the
steady-state distribution of the ordinary random walk on the graph,
governed by the transition matrix $Q$: $\pi Q = \pi$ (see Appendix
\ref{rw_stat_dist}).

The ratio $-\lambda_k^0/\lambda_k^1$ in Eq. (\ref{ps_first2}) is the
pole of the approximate Laplace-transformed propagator and thus an
approximation of the real pole $s_k$.  This approximation can only be
valid for poles with the small absolute value $|s_k|$.  Denoting
$\tau_m = \max\limits_{k>0}\{ \lambda_k^1/\lambda_k^0\} =
\lambda_{k_m}^1/\lambda_{k_m}^0$ (for some index $k_m$) as the largest
time scale, we get the long-time exponential approach to the
steady-state distribution:
\begin{align}
P_{x_0x}(t) \simeq p^{\rm st}_x  + t_x \frac{v_{k_m}^0(x_0) \, u_{k_m}^0(x)}{\lambda_{k_m}^1} \, e^{- t/\tau_m} .
\label{ps_first3}
\end{align}

The above analysis is not applicable when at least one mean travel
time is infinite.  We sketch the main steps of the asymptotic analysis
for the particular situation when all probability densities
$\psi_{xx'}(t)$ exhibit heavy tails with {\it the same} scaling
exponent $0< \alpha < 1$: $\psi_{xx'}(t) \propto t^{-1-\alpha}$ or,
equivalently,
\begin{align}
\tilde{\psi}_{xx'}(s) = 1 - s^\alpha \tau_{xx'}^\alpha + o(s^\alpha), 
\end{align}
with possibly different time scales $\tau_{xx'}$.  The propagator is
then approximated as
\begin{align}
\tilde{P}_{x_0x}(s) \simeq s^{\alpha-1} 
t_x \bigg[(I-Q + s^\alpha T)^{-1}\bigg]_{x_0x},
\label{prop_special_case2}
\end{align}
with the matrix $T$ being still defined by Eq.~(\ref{eq:T}), in which
$\langle \T_{xx'}\rangle$ are replaced by $\tau_{xx'}^\alpha$, and
$t_x$ is defined by Eq.~(\ref{eq:tx}).

For small $s$, one can apply the same perturbation theory, in which
$s$ is replaced by $s^\alpha$, to get
\begin{align}
\tilde{P}_{x_0x}(s)  \simeq s^{\alpha-1} t_x \sum_{k\geq 0} \frac{v_{k}^0(x_0) \, u_{k}^0(x)}{\lambda_{k}^0 + s^\alpha \lambda_{k}^1}  .
\label{ps_first4}
\end{align}
The formal inversion of the Laplace transform yields the long-time
asymptotic approach to the steady-state
\begin{align}
P_{x_0x}(t) &\simeq p^{\rm st}_x + 
 t_x  \frac{v_{k_m}^0(x_0) \, u_{k_m}^0(x)}{\lambda_{k_m}^1} E_{\alpha}(- t^\alpha /\tau_m) ,
\label{ps_first5}
\end{align}
where $E_\alpha(z)$ is the Mittag-Leffler function.  Since the
Mittag-Leffler function exhibits a slow, power law decay,
$E_\alpha(-z) \simeq z^{-1}/\Gamma(1-\alpha)$ as $z\to \infty$, the
major difference with the former case of finite mean travel times is a
much slower approach to the steady-state limit, which is caused by
long traps.

\section{Discussion}
\label{Discus_sec}

The HCTRW model naturally describes multi-state systems, for which the
states are represented by nodes and the probability densities
$\psi_{xx'}(t)$ characterize the inverse of random exchange rates.
The simplest example is a two-state system, which switches randomly
between two states $1$ and $2$ after random times $\T_{12}$ and
$\T_{21}$ drawn from the probability densities $\psi_{12}(t)$ and
$\psi_{21}(t)$.  In this case, $Q = \biggl(\begin{array}{c c} 0 & 1 \\
1 & 0 \\ \end{array}\biggr)$, and the Laplace-transformed propagator
reads as
\begin{eqnarray}  \nonumber
&& \tilde{P}(s) = \frac{1}{s(1 - \tilde{\psi}_{12}(s) \tilde{\psi}_{21}(s))} \\
&& \times \left(\begin{array}{c c}
1 - \tilde{\psi}_{12}(s) & (1-\tilde{\psi}_{21}(s)) \tilde{\psi}_{12}(s) \\
(1-\tilde{\psi}_{12}(s)) \tilde{\psi}_{21}(s) & 1 - \tilde{\psi}_{21}(s) \\ \end{array} \right),
\end{eqnarray}
where $\tilde{P}(s)$ is a $2\times 2$ matrix notation for
$\tilde{P}_{x_0x}(s)$.  In particular, the non-Markovian dynamics of
such a simple system was established when $\psi_{xx'}(t)$ are not
exponential densities \cite{Boguna2000}.  The HCTRW formalism
naturally extends this analysis to a multi-state system, which
randomly switches between $N$ different states.  More generally, the
HCTRW framework can be related to the theory of renewal processes
\cite{WilmerLevin}, to random walks in random environments
\cite{Hughes,Zeitouni2002}, and to persistent CTRW \cite{Jaume2017}.

The matrix $H = I - \tilde{Q}(s)$ can be seen as a normalized form of
a weighted discrete Laplacian on a graph, which is also related to the
model of random resistor networks \cite{Mieghem}.  Since the
generalized transition matrix $\tilde{Q}(s)$ couples the structure of
the graph (the matrix $Q$) to the spatio-temporal dynamics of the
walker on that graph (the densities $\psi_{xx'}(t)$), it is natural to
distinguish the effects of both aspects.  In particular, one can
investigate how the spectral properties of the matrix $\tilde{Q}(s)$
are affected by \emph{structural} (or geometric) and
\emph{spatio-temporal} (or distributional) perturbations.  In the
former case, one changes the structure of the graph (e.g., by adding,
removing, or modifying some links).  In the latter case, the graph is
kept fixed but the densities $\phi_{xx'}(t)$ are modified.  Analytical
estimates for the propagator under spatio-temporal perturbations can
be derived by using the time-dependent perturbation theory
\cite{Griffiths}, and approximations for the smallest eigenvalue
\cite{Zou2009,Hunter1986}.

\subsection{Absorbing boundary, bulk reactions, and first passage phenomena}

In contrast to continuous-space problems, one can naturally
accommodate boundary conditions through the stochastic matrix $Q$,
with no change to the HCTRW formalism.  In fact, a reflecting boundary
is intrinsically implemented by the mere fact of a finite-size matrix
$Q$.  An absorbing boundary or a target can be implemented by adding a
``sink site'' $x^*$ to the graph, such that $Q_{x^*x^*} = 1$, i.e.,
any particle that comes to $x^*$ remains trapped at this site.  The
geometric structure of the absorbing boundary (or the target) is
captured through the elements of the matrix $Q_{xx^*}$, i.e., the
probabilities of arriving at the sink site from other sites of the
graph.  The propagator $P_{x_0x}(t)$ is then interpreted as the
probability for a walker started at $x_0$ to be at a site $x$ at time
$t$ {\it without being absorbed}.  In turn, $P_{x_0x^*}(t)$ is the
probability of being absorbed by time $t$, whereas $S_{x_0}(t) = 1 -
P_{x_0x^*}(t)$ is the survival probability.  As a consequence,
$P_{x_0x^*}(t)$ can be interpreted as the cumulative probability
distribution of the first passage time (FPT) to the sink site (or to
the absorbing boundary), whereas $\rho_{x_0}(t) = \partial
P_{x_0x^*}(t)/\partial t$ is the probability density of this FPT.  The
mean FPT is simply $\tilde{P}_{x_0x^*}(0)$, and other moments of the
FPT are expressed as derivatives of the Laplace-transformed propagator
$\tilde{P}_{x_0x^*}(s)$ at $s=0$.  One can also easily treat partially
absorbing boundaries 
\cite{Collins49,Sano79,Sapoval94,Grebenkov03,Grebenkov06,Grebenkov07,Singer08,Grebenkov10b,Rojo12}
by allowing nonzero leakage probability from the sink site $x^*$.

If a particle can disappear or loose its activity during diffusion,
FPT problems for such ``mortal'' walkers
\cite{Abad10,Abad12,Abad13,Yuste13,Abad15,Meerson15a,Meerson15b,Grebenkov17}
can be treated by introducing two sink sites, $x^*_1$ and $x^*_2$,
that represent an absorbing boundary and a reactive bulk.  Using the
exchange time distributions $\psi_{xx^*_2}(t)$ depending on $x$, one
can model space-dependent bulk reaction rates.  Note also that
$P_{x_0x^*_1}(\infty)$ is the splitting probability, i.e., the
probability of the arrival on $x^*_1$ before arriving on $x^*_2$
(i.e., the arrival to the target before dying or loosing activity).
If there are many sink sites $x^*_1,\ldots,x^*_k$,
$P_{x_0x^*_i}(\infty)$ are the hitting probabilities (a discrete
analog of the harmonic measure).

All these conventional concepts of first passage phenomena
\cite{Redner2002} are accessible through the mathematical formalism of
HCTRW which plays thus a unifying role.  The main advantage of this
approach is the reduction of the sophisticated {\it dynamics} in
heterogeneous media to the spectral properties of the governing matrix
$\tilde{Q}(s)$ which generalizes the stochastic matrix $Q$.  In the
same way as the structural features of the medium that are relevant
for simple random walks were captured through the spectral properties
of the transition matrix $Q$ \cite{Lin13}, the spatio-temporal
heterogeneities of the medium are captured by the spectral properties
of the generalized transition matrix $\tilde{Q}(s)$.

\subsection{Effects of spatio-temporal heterogeneities on first passage times}

If one is primarily interested in the impact of spatio-temporal
heterogeneities onto the diffusive dynamics, one can choose the
simplest geometric setting, a discretized interval, represented by a
graph with $N = 100$ sites.  We consider a symmetric HCTRW on this
graph, with equal probabilities to move to the left and to the right.
The nodes $x^* = 1$ and $x = 100$ are respectively absorbing and
reflecting.  This fixes the transition matrix $Q$ as follows: $Q_{xx'}
= \frac12 \delta_{x,x'-1} + \frac12 \delta_{x,x'+1}$ for $1 < x < N$;
$Q_{1x'} = \delta_{1,x'}$; and $Q_{Nx'} = \frac12 \delta_{N,x'-1} +
\frac12 \delta_{N,x'}$.  In turn, the temporal aspects of diffusion,
represented by travel time densities $\psi_{xx'}(t)$, will be
explored.  Note that the results do not depend on the choice of the
density $\psi_{x^*x^*}(t)$ at the sink site (see Appendix
\ref{sec:no_dependence}).

\begin{figure}	
\includegraphics[width=85mm]{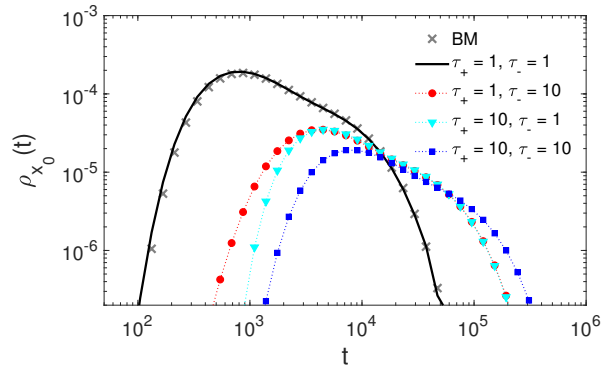}
\caption{
FPT probability density $\rho_{x_0}(t)$ for symmetric HCTRW on a
discrete interval with $N = 100$ sites, the absorbing endpoint at $x^*
= 1$, and the reflecting endpoint at $x=100$.  We set $x_0 = 50$ and
$\tilde{\psi}_{xx'}(s) = (1 + s\tau_{xx'})^{-1}$, with $\tau_{xx'} =
\tau_{+} \delta_{x,x'-1} + \tau_{-} \delta_{x,x'+1}$.  Gray crosses
show the density $\rho_{z_0}^{\rm BM}(t)$ for Brownian motion on the
unit interval, with $D = 1/(2N^2)$ and $z_0 = x_0/N$.}
\label{fig_FPTD_theor_homo}
\end{figure}

Although various diffusive characteristics are available, we focus on
the probability density $\rho_{x_0}(t)$ of the FPT.  For each
considered example, we compute this density by using the Talbot
algorithm \cite{Talbot79} for a numerical inversion of the Laplace
transform of $s \tilde{P}_{x_0x^*}(s)$ (with $x^* =1$).  To validate
this inversion procedure, we compare $\rho_{x_0}(t)$ in the
homogeneous case with $\tilde{\psi}_{xx'}(s) = (1 + s\tau)^{-1}$, to
the known solution of the FPT probability density for Brownian motion
on the unit interval $(0,1)$ with absorbing (resp. reflecting)
endpoint at $0$ (resp., at $1$):
\begin{equation}
\rho_{z_0}^{\rm BM}(t) = \pi D \sum\limits_{n=0}^\infty (n+1/2) \sin(\pi (n+1/2) z_0) e^{-\pi^2 (n+\frac12)^2 Dt} ,
\end{equation}
where $D$ is the diffusion coefficient, and $z_0 = x_0/N$.  Setting $D
= a^2/(2\tau)$ with $a = 1/N$ being the inter-site distance, one
expects that the homogeneous diffusion on this graph is a discrete
approximation of Brownian motion so that $\rho_{x_0}(t)$ and
$\rho_{x_0/N}^{\rm BM}(t)$ are close to each other.  One can see an
excellent agreement between two functions (shown by solid line and
crosses) in Fig.~\ref{fig_FPTD_theor_homo}, except at short times at
which small deviations can be attributed to the discretization of the
interval by $N$ points.  After this validation, we will reveal the
impact of spatio-temporal heterogeneities by comparing all results to
the homogeneous case, with $\tilde{\psi}_{xx'}(s) = (1 + s\tau)^{-1}$.

First, we illustrate the effect of nonsymmetric travel times.  For
this purpose, we set $\tilde{\psi}_{xx'}(s) = (1 + s\tau_{xx'})^{-1}$
with $\tau_{xx'} = \tau_{+} \delta_{x,x'-1} + \tau_{-}
\delta_{x,x'+1}$.  In other words, we consider a random walker jumping
with exponentially distributed travel times but the mean time to jump
to the left, $\tau_-$, is different from the mean time to jump to the
right, $\tau_+$.  This difference may originate, e.g., from a
potential inside channels connecting neighboring pores.  We emphasize
that the probabilities of jumping to the left and to the right remain
equal.  Figure~\ref{fig_FPTD_theor_homo} compares four cases: (a)
$\tau_{+} = \tau_{-} = 1$; (b) $\tau_{+} = 10$, $\tau_{-} = 1$; (c)
$\tau_{+} = 1$, $\tau_{-} = 10$; and (d) $\tau_{+} = \tau_{-} = 10$.
As expected, the cases (a) and (d) yield the fastest and the slowest
arrival to the sink site, whereas the cases (b) and (c) stand in
between.  Note that the cases (b) and (c) exhibit the identical
behavior at long times when the walker performs many jumps and the
asymmetry between jumps to the left and to the right is averaged out.
In turn, there is a notable difference at short times: when the number
of jumps is not large, it matters whether the travel time to the left
(towards the sink) is small or large.

\begin{figure}	
\includegraphics[width=85mm]{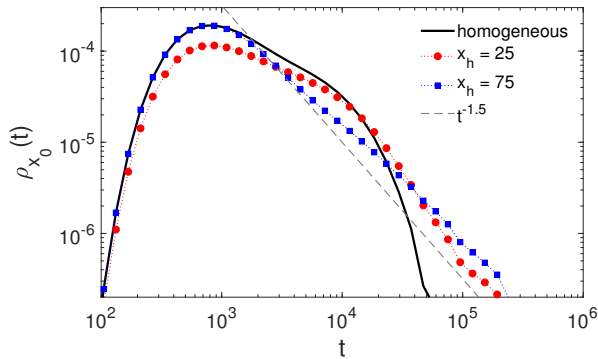}
\caption{
FPT probability density $\rho_{x_0}(t)$ for symmetric HCTRW on a
discrete interval with $N = 100$ sites, the absorbing endpoint at $x^*
= 1$, and the reflecting endpoint at $x=100$.  We set $x_0 = 50$,
$\tau = 1$, and $\tilde{\psi}_{xx'}(s) = (1 + s\tau)^{-1}$ for all
$x$, except for a trapping site at $x_h$ for which
$\tilde{\psi}_{x_hx'}(s) = (1 + (s\tau)^\alpha)^{-1}$, with $\alpha =
0.5$.  Two cases $x_h = 25$ and $x_h = 75$ are compared to the
homogeneous case without trapping site (solid line).  Dashed line
shows a power law decay $t^{-1-\alpha}$.  }
\label{fig_FPTD_theor_het_point}
\end{figure}

Second, we demonstrate the effect of adding a single trapping site
with reversible binding kinetics.  For this purpose, we consider the
homogeneous interval with $\tilde{\psi}_{xx'}(s) = (1 + s
\tau)^{-1}$, except for one point $x_h$, at which
$\tilde{\psi}_{x_hx'}(s) = (1 + (s \tau)^\alpha)^{-1}$, with a scaling
exponent $\alpha = 0.5$.  This corresponds to the Mittag-Leffler
distribution of exchange times.  Since the mean waiting time at the
trapping site is infinite, a random walker can remain trapped much
longer at this particular site, as compared to other sites.  Figure
\ref{fig_FPTD_theor_het_point} shows the probability density
$\rho_{x_0}(t)$ with $x_0 = 50$ for three cases: no trapping site (the
reference case), trapping site at $x_h = 25$ and trapping site at $x_h
= 75$.  The two latter cases are qualitatively different because the
walker is always trapped at $x_h = 25$ on the way to the sink at $x^*
= 1$, whereas the trapping site at $x_h = 75$ may be not be visited
when started at $x_0 = 50$.  In the latter case, the density
$\rho_{x_0}(t)$ coincides with that for the homogeneous case at short
times because the short trajectories to the sink do not pass through
the trapping site at $x_h = 75$.  In turn, significant deviations
appear at long times.  Indeed, eventual traps with the infinite mean
trapping time drastically changes the propagator so that the density
$\rho_{x_0}(t)$ exhibits a slow, power law long-time asymptotic decay:
$\rho_{x_0}(t) \propto t^{-1-\alpha}$, in analogy to
Eq. (\ref{ps_first5}).  In particular, the mean FPT to the sink is
infinite, regardless the position of the trap.

\begin{figure}	
\includegraphics[width=85mm]{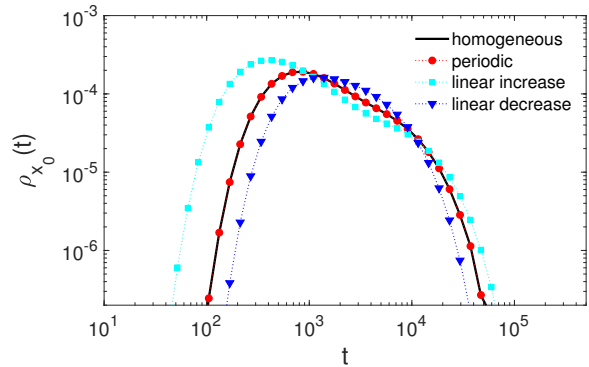}
\caption{
FPT probability density $\rho_{x_0}(t)$ for symmetric HCTRW on a
discrete interval with $N = 100$ sites, the absorbing endpoint at $x^*
= 1$, and the reflecting endpoint at $x=100$.  We set $x_0 = 100$,
$\tau = 1$, and $\tilde{\psi}_{xx'}(s) = (1 + s\tau_x)^{-1}$, with
four choices: $\tau_x = \tau$ (homogeneous case), $\tau_x = \tau(1 +
0.5 \sin(4\pi x/N))$, $\tau_x = ax$ (linear increase), $\tau_x =
a(N+1-x)$ (linear decrease), with $a = 2\tau/(N(N+1))$. }
\label{fig_FPTD_theor_hetero}
\end{figure}

Third, we look at the effect of spatial variations of the mean travel
time $\tau_x$ by setting $\tilde{\psi}_{xx'}(s) = (1 + s
\tau_x)^{-1}$.  Such a HCTRW can be viewed as a microscopic model
of heterogeneous diffusion processes with space-dependent diffusion
coefficient \cite{Cherstvy13,Pineda2014} which can also mimic crowding
effects \cite{Ghosh16}.  We consider four choices for $\tau_x$: a
constant, $\tau_x = \tau$ (the reference case); a periodic variation,
$\tau_x = \tau(1 + c \sin(q\pi x/N))$; a linear growth from the sink,
$\tau_x = ax$; and a linear decrease towards the sink, $\tau_x = a(N+1
- x)$.  For a proper comparison of these cases, we choose the
functions $\tau_x$ to have the same mean travel time over the
interval.  We use the arithmetic mean: $\frac{1}{N}
\sum\nolimits_{x=1}^N \tau_x = \tau$, as justified below.  In
particular, we set $a = \frac{2\tau}{N(N+1)}$ and take $q$ to be an
integer (one also imposes $|c| < 1$ to ensure the positivity of
$\tau_x$).  Figure \ref{fig_FPTD_theor_hetero} shows the probability
density $\rho_{x_0}(t)$ at $x_0 = 100$ for these cases.  As expected,
periodic variations of the mean travel time have no effect on the
first passage time, in comparison to the homogeneous case.  In fact,
these variations are averaged out by passing through all the sites.
This observation justifies our choice of using the arithmetic mean:
the first passage time can be viewed as a weighted sum of travel times
between visited sites.  We also checked that the probability density
$\rho_{x_0}(t)$ does not depend on the amplitude $c$ and the frequency
$q$ for a broad range of these parameters (not shown).  In turn, if
the starting point $x_0$ is not set at a site with $\tau_{x_0} =
\tau$ (here, at $x_0 = 100$), then differences between the homogeneous
and periodic cases can emerge.  For instance, if $q = 1$, $c > 0$, and
$x_0 = 25$, then a walker would on average take longer travel times,
as $\tau_x > \tau$ for $x$ between $1$ and $50$.  This difference is
particularly important at short times.

Now we turn to the linear dependence of $\tau_x$ on $x$.  The spatial
heterogeneity of travel times strongly affects the probability density
$\rho_{x_0}(t)$: the distribution of FPT is much wider in the case
when the mean travel time $\tau_x$ increases from the sink site,
$\tau_x = a x$, as compared to the case of decreasing $\tau_x =
a(N+1-x)$.  Indeed, the probability density $\rho_{x_0}(t)$ at long
times is determined by long trajectories, which stayed away from the
sink.  Since the random walker samples preferentially the sites far
from the sink, the FPT to the sink is longer in the case of linearly
increasing $\tau_x$ and shorter in the case of linearly decreasing
$\tau_x$.  The argument is inverted at short times when the density
$\rho_{x_0}(t)$ is determined by short trajectories when the walker
moves preferentially towards the sink.

\section{Conclusions}
\label{Concluss_sec}

We presented a new model of heterogeneous continuous time random
walks, which generalizes CTRW by allowing a heterogeneous distribution
of travel times between sites.  This model merges two important and
rapidly developing research directions: continuous-time random walks
as a generic model of anomalous transport, and discrete-time random
walks on graphs and networks.  We derived the analytical formula
(\ref{hctrw_final}) for the HCTRW propagator in the Laplace domain and
discussed its inversion to time domain.  In particular, the
perturbative analysis of the matrix $I - \tilde{Q}(s)$ yields the
long-time asymptotic behavior.  More generally, the complex diffusive
dynamics in multiscale structures with spatio-temporal heterogeneities
was related to the spectral properties of the generalized transition
matrix $\tilde{Q}(s)$.  In this light, a rigorous extension of this
study to infinite graphs (or, equivalently, the limit of increasing
graphs) presents an important perspective.  In this situation, the
steady-state distribution may not exist (as for a simple random walk
on an infinite lattice), whereas the spectrum of the generalized
stochastic matrix may be continuous.  The derivation of a macroscopic
description of HCTRW on very large (or infinite) graphs
\cite{Asz1993}, like fractional diffusion equation for CTRW, remains
an open problem.  This analysis can shed a light onto space-dependent
diffusion equations and provide their microscopic models.

In order to reveal the effects of spatio-temporal heterogeneities onto
the diffusive dynamics, we kept the geometric structure as simple as
possible.  The next step consists in coupling these heterogeneities to
the structural complexity of graphs and networks
\cite{Albert02,Noh04,Colizza07,Sood07,Haynes09,Nicolaides10,Barthelemy11,Holme12,Perra12,Hwang12,Skarpalezos13,Goutsias13,Kozak14,Bovaventura14,Agliari16}.
For instance, one can study HCTRW on some fractal trees and networks,
for which the spectral properties are relatively well known
\cite{Grabow12,Julati2013,Sole-Ribalta13}.  Even a simple random walk
on fractal structures such as tree graphs, in combination with the
particular distributions of waiting times, often leads to anomalous
diffusion \cite{Tamm2015,Spanner16}.  Since coarse-graining methods
have been extensively developed over the last decade
\cite{Vocka00,Picard07,Klimenko12,Varloteaux13}, the HCTRW framework
has a promising application for studying transport properties in
porous materials.  Other potential applications include transportation
systems (with the intricate interrelation between the traffic and the
complex topology of the roads graph or airflight connections),
electric networks, as well as internet and social networks
\cite{Barrat04,Noulas15}.

\begin{acknowledgments}
The authors acknowledge the support under Grant
No. ANR-13-JSV5-0006-01 of the French National Research Agency.
\end{acknowledgments}

\appendix
\section{Basic properties and technical relations}

\subsection{Reduction to Montroll-Weiss formula}
\label{sec:Montroll}

We show that HCTRW framework leads to the Montroll-Weiss formula for
CTRW on a lattice \cite{Montroll1973}.  The spatial and temporal
components of the waiting time distribution of CTRW are separated:
\begin{align}
Q_{x'x}(t) = Q_{x'x} \psi(t),
\end{align}
where $\psi(t)$ is the waiting time probability density.  Then
Eq.~(\ref{hctrw_final}) becomes
\begin{align}
\tilde{P}_{x_0x}(s)= \frac{1-\tilde{\psi}(s)}{s}[(I-Q\tilde{\psi}(s))^{-1}]_{x_0x}.
\label{ps_special}
\end{align}
For one-dimensional lattice $\Z$, the discrete Fourier transform yields
\begin{align}
{\mathcal F}_k\{\tilde{P}_{x_0x}(s) \} & =  \sum\limits_{x=\infty}^{\infty} \tilde{P}_{x_0x}(s)e^{ikx} \\
& = e^{ikx_0}\frac{1-\tilde{\psi}(s)}{s} \frac{1}{1-\tilde{\psi}(s)(1-\lambda_{k}^0)} .
\end{align}
This is the Montroll-Weiss formula for CTRW in Laplace-Fourier domain,
where $1-\lambda_{k}^0 = qe^{-ik} + (1-q)e^{ik}$ is the characteristic
function of the jump distribution on the lattice, with probability $q$
(resp., $1-q$) to jump to the left (resp., to the right).  The
calculation extends to $\Z^d$ with $d$-dimensional discrete Fourier
transform.

\subsection{Spectrum of the generalized transition matrix}
\label{sec:Qnorm}

Since $\tilde{\psi}_{xx'}(s)$ is the Laplace transform of a
probability density of a positive random variable, one has
$\tilde{\psi}_{xx'}(s) \geq 0$, $\tilde{\psi}_{xx'}(0)=1$, and
\begin{equation*}
\tilde{\psi}_{xx'}(s+\delta) - \tilde{\psi}_{xx'}(s) = \int_{0}^{\infty}\psi_{xx'}(t) e^{-ts}(e^{-\delta t}-1) dt < 0 
\end{equation*}
for any $\delta>0$ (we excluded the trivial distribution with
$\psi_{xx'}(t) = \delta(t)$, for which the integral is equal to $0$).
As a consequence, $\tilde{\psi}_{xx'}(s)$ is a monotonously decreasing
function on $(0,\infty)$, and thus
\begin{equation}  \label{eq:tilde_psi_ineq}
0\leq \tilde{\psi}_{xx'}(s) < 1 \qquad (s > 0).
\end{equation}

The matrix $\tilde{Q}(s)$ is a real (nonsymmetric) matrix with
nonnegative elements.  We note that the matrix $\tilde{Q}(s)$ is not
necessarily irreducible that allows us to consider, e.g., sink sites.
According to the Perron-Frobenius theorem for nonnegative matrices,
there exists a nonnegative eigenvalue $\lambda_0$ such that the
corresponding eigenvector $v_0$ has nonnegative components, and the
other eigenvalues $\lambda_k$ are bounded in the absolute value:
$|\lambda_k| \leq \lambda_0$.  Since the sum of the elements of the
matrix $\tilde{Q}_{xx'}(s)$ in each column does not exceed $1$, one
gets $\lambda_0 \leq 1$.  In fact, denoting $v_0(x)$ the maximal
component of the vector $v_0$, $v_0(x) =
\max\limits_{x'} \{v_0(x')\} > 0$, one has
\begin{equation*}
\lambda_0 v_0(x) = \sum\limits_{x'} \tilde{Q}_{xx'} v_0(x') \leq \max\limits_{x'} \{v_0(x')\} \sum\limits_{x'} \tilde{Q}_{xx'} \leq v_0(x)
\end{equation*}
that implies $\lambda_0 \leq 1$.  Moreover, the inequality is strict
for $s > 0$ due to Eq. (\ref{eq:tilde_psi_ineq}).  As a consequence,
the matrix $I-\tilde Q(s)$ is invertible for any $s>0$.

\subsection{Normalization of the HCTRW propagator}
\label{normal_hctrw}

We check the normalization of the HCTRW propagator.  From
Eq.~(\ref{hctrw_final}) we get:
\begin{align}
&\sum_{x} \tilde{P}_{x_0x }(s)= \\ 	\nonumber
&\frac{1}{s}\sum_x  (1-\sum_{x'} \tilde{Q}(s)_{xx'})
[(I-\tilde{Q}(s))^{-1}]_{x_0x}= 
\\ \nonumber 
&\frac{1}{s}\sum_{x,x'} \bigg[( I-\tilde{Q}(s))^{-1}\bigg]_{x_0x}
\bigg[(I-\tilde{Q}(s))\bigg]_{xx'} = \frac{1}{s}.
\end{align}
So that $\sum_{x}P_{x_0x}(t) =1$ for any $t$ and $x_0$.

The normalization of the approximate expression
Eq.~(\ref{prop_special_case}) is also fulfilled:
\begin{align}
&\sum_x (\sum_{x'} T_{xx'})
[(I-Q + sT)^{-1}]_{x_0x} =  \\ \nonumber 
&\sum_{x,x'}[(I-Q+sT)^{-1}]_{x_0x}T_{xx'} = \\ \nonumber 
&\sum_{x'} [(I-Q+sT)^{-1}T]_{x_0x'} = 
\\ \nonumber 
&\frac{1}{s}\sum_{x'}[(I-Q+sT)^{-1}(sT+I-Q - (I-Q))]_{x_0x'} = \\ \nonumber 
&\frac{1}{s}\bigg(1 -\sum_{x,x'}[(I-Q+sT)^{-1}]_{x_0x}[I-Q]_{xx'}\bigg) =
\frac{1}{s},
\end{align}
where the last implication is valid because $\sum_{x'}[(I-Q)]_{xx'} =
0$, independently of $x$ and $s$.

\subsection{Stationary distribution of a simple random walk on a graph}
\label{rw_stat_dist}

We recall the basic result about the stationary distribution of a
simple random walk on a graph $G = (V,E)$ with $E$ for the set of
edges.  In this model, the transition probability from $x'$ to $x$ is
$Q_{x'x} = 1/\deg_{x'}$, where $\deg_{x'}$ denotes the number of edges
incident with the node $x'$.  The stationary distribution is $\pi_x =
\deg_x/(2|E|)$, where $|E|$ is the number of edges.  In fact, we have
\begin{align}
\sum_{x'} \pi_{x'} Q_{x'x} = \sum_{x' \in {\mathcal A}(x)} \frac{\deg_{x'}}{2|E|} \, \frac{1}{\deg_{x'}} = \frac{\deg_{x}}{2|E|} = \pi_x,
\end{align}
where the sum runs over all sites $x'$ adjacent to $x$ (for other
sites $Q_{x'x}$ is zero).  We get thus $\pi Q = \pi$.

\subsection{No dependence on $\psi_{x^*x^*}(t)$}
\label{sec:no_dependence}

As intuitively expected, the propagator in the presence of a sink at
$x^*$ does depend on the choice of the corresponding travel time
probability density $\psi_{x^*x^*}(t)$.  For simplicity of notations,
let $x^* = 1$ so that the governing matrix has the form
\begin{equation}
H = \left(\begin{array}{c c c c} 
\phi & 0 & 0 & \ldots  \\
 & & & \\
& & \hat{H} & \\
 & & & \\  \end{array} \right),
\end{equation}
where $\phi = 1 - \tilde{\psi}_{x^*x^*}(s)$ and $\hat{H}$ is the
remaining matrix of size $(N-1)\times N$.  The elements of the inverse
of $H$ can be formally written in terms of minors as
\begin{equation} \label{eq:tildeP_minors}
\tilde{P}_{x_0x}(s) = \frac{1 - \sum\nolimits_{x'} \tilde{Q}_{xx'}(s)}{s} \, 
\frac{(-1)^{x_0+x} \det(\M_{xx_0}(H))}{\det(H)} ,
\end{equation}
where $\M_{xx_0}(H)$ is the matrix obtained from $H$ by removing the
row $x$ and the column $x_0$.  We consider separately two cases: $x
\ne x^*$ and $x = x^*$:

(i) In the former case, the first factor in
Eq. (\ref{eq:tildeP_minors}) does not contain
$\tilde{\psi}_{x^*x^*}(s)$.  Using the Laplace's formula and the
structure of the matrix $H$, one gets $\det(H) = \phi ~
\det(\M_{x^*x^*}(H))$.  Similarly, $\det(\M_{xx_0}(H)) = \phi ~
\det(\M_{x^*x^*}(\M_{xx_0}(H)))$ so that the factor $\phi$ containing
$\tilde{\psi}_{x^*x^*}(s)$ is canceled, whereas the remaining minors
do not contain $\phi$.

(ii) In the case $x = x^*$, Eq. (\ref{eq:tildeP_minors}) becomes
\begin{equation}
\tilde{P}_{x_0x^*}(s) = \frac{\phi}{s} \, 
\frac{(-1)^{x_0+x} \det(\M_{x^*x_0}(H))}{\phi~ \det(\M_{x^*x^*}(H))} ,
\end{equation}
where we used that $\tilde{Q}_{x^*x'} = \delta_{x^*x'}
\tilde{\psi}_{x^*x^*}(s)$.  Once again, the factor $\phi$
is canceled whereas the minors $\M_{x^*x_0}(H)$ and $\M_{x^*x^*}(H)$
do not contain $\phi$.  We conclude that the propagator
$\tilde{P}_{x_0x}(s)$ does not depend on $\tilde{\psi}_{x^*x^*}(s)$.

\section{Explicit solutions for circular graphs}
\label{hctrw_circ} 

In general, a numerical Laplace inversion is needed to get the HCTRW
propagator in time domain.  Here we provide an example when the
inversion can be performed explicitly.

\begin{figure}
\includegraphics[width=80mm]{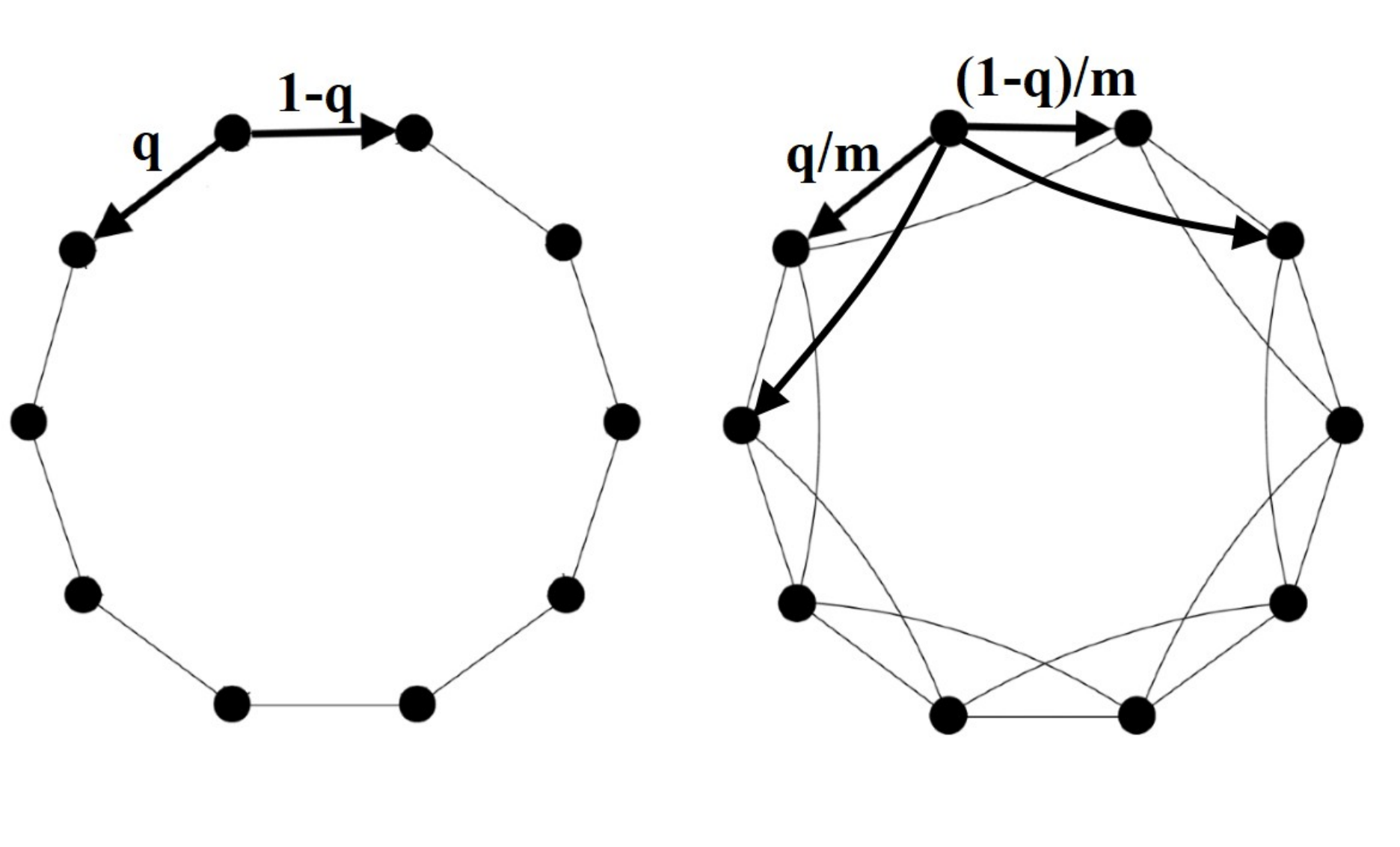}  
\caption{
$m$-circular graph with $N = 10$ nodes is shown for $m=2$ (left) and
$m=4$ (right).  Arrows indicate possible jumps to neighboring nodes. }
\label{m_circ}
\end{figure}

Let us consider the asymmetric random walk on a $m$-circular graph
(also known as a regular small-world graph) with $N$ nodes, where the
degree $m$ of each node is even (Fig.~\ref{m_circ}).  Such graphs have
a circular transition matrix $Q$ with $m$ non-zero elements in each
row \cite{Mieghem}.  Let us set transition probabilities for each node
to be $2q/m$ (jumps to the ``right'') and $2(1-q)/m$ (jumps to the
``left'').  We also choose $\tilde{\psi}_{xx'}(s)$ to be equal to
$\tilde{\psi}_+(s)$ for jumps to the right and to $\tilde{\psi}_-(s)$
for jumps to the left.  The components of an eigenvector of the
circular matrix $I-\tilde{Q}(s)$ are given by:
\begin{align}
v_k(x) & = e^{2\pi ik x/N}/\sqrt{N},
\end{align}
where $k = 0,1,\ldots,N-1$ and $x = 1,\ldots,N$.  Denoting $\gamma =
e^{2\pi i /N}$, the eigenvalues $\lambda_k(s)$ of $I-\tilde{Q}(s)$
are:
\begin{align} \label{lambda_k_s}
\lambda_k(s) & = 1 - q \tilde{\psi}_-(s) \gamma_k  - (1-q) \tilde{\psi}_+(s) \gamma_{-k}   ,
\end{align}
where
\begin{equation}
\gamma_k = \frac{2}{m} \, \frac{1 - e^{2\pi i km/(2N)}}{e^{-2\pi ik/N}-1} .
\end{equation}
These spectral quantities fully determine the propagator in the
Laplace domain according to Eq.~(\ref{hctrw_final}).  

In the particular case of exponential distributions
$\tilde{\psi}_{\pm}(s) = (1 + s\tau_{\pm})^{-1}$, one easily gets the
explicit form of the propagator in time domain.  For this purpose we
represent $\lambda_k(s)$ as
\begin{align}
&\lambda_k(s) = \frac{\tau_+ \tau_- s^2 + B_{k}s + C_{k} }{(1+s\tau_+)(1+s\tau_-)},
\label{root_m_circ}
\end{align}
where  
\begin{eqnarray*}
B_{k} &=& (\tau_+ + \tau_-) - \tau_+ q \gamma_k - \tau_- (1-q) \gamma_{-k} , \\
C_{k} &=& 1 - q \gamma_k - (1-q) \gamma_{-k} .
\end{eqnarray*}
Since $C_{0} = 0$, one has 
\begin{align}
\lambda_0(s) = \frac{\tau_+ \tau_- s (s + {\omega})}{(1+s\tau_+)(1+s\tau_-)} ,
\end{align}
where ${\omega} = (1-q)/\tau_- + q/\tau_+$.  As a consequence, the
Laplace transform of the propagator is
\begin{align}
\tilde{P}_{x_0x}(s) & = \frac{1}{sN} + \frac{1}{N} 
\sum\limits_{k=1}^{N-1} \frac{e^{2\pi ik(x-x_0)/N}}{{s}_{k}^+ - {s}_{k}^-}  \\ \nonumber
&\times \biggl( \frac{{\omega} + {s}_{k}^+}{s - {s}_{k}^+} - \frac{{\omega} + {s}_{k}^-}{s - {s}_{k}^-}\biggr),
\label{psx0x_mcirc}
\end{align}
where 
\begin{equation*}
s_k^{\pm} = \frac{-B_{k} \pm \sqrt{B_{k}^2 - 4\tau_+ \tau_- C_{k}}}{2\tau_+ \tau_-} .
\end{equation*}
The Laplace inversion yields the propagator in time domain:
\begin{align}
P_{x_0x}(t) & = \frac{1}{N} + \frac{1}{N} \sum_{k=1}^{N-1} \frac{e^{2\pi k(x-x_0)/N}}{s^+_{k}-s^-_{k}} \\ \nonumber
&\times \biggl((\omega + s^+_{k})e^{s^+_{k}t} - (\omega + s^-_{k})e^{s^-_{k}t}\biggr).
\end{align}

In the case of Mittag-Leffler distribution of travel times,
$\tilde{\psi}_\pm(s) = 1/(1+s^{\alpha}\tau_{\pm}^{\alpha})$, one can
simply replace $s$ by $s^\alpha$ and $\tau_\pm$ by $\tau_\pm^\alpha$
in the above expressions for $\lambda_k(s)$, $B_k$, $\omega$, and
$s_k^\pm$.  As a consequence, the propagator in time domain reads
\begin{align}
& P_{x_0x}(t) = \frac{1}{N} + \frac{1}{N} \sum_{k=1}^{N-1} \frac{e^{2\pi k(x-x_0)/N}}{s^+_{k}-s^-_{k}} \\ \nonumber
&\times \biggl((\omega + s^+_{k}) E_{\alpha}(s^+_{k}t^\alpha)  - (\omega + s^-_{k}) E_{\alpha}(s^-_{k}t^\alpha)\biggr),
\end{align}
where $E_\alpha(z)$ is the Mittag-Leffler function.

\end{document}